\input harvmac.tex
\input epsf.tex
\parindent=0pt
\parskip=5pt

\hyphenation{satisfying}

\def\IR{{\hbox{{\rm I}\kern-.2em\hbox{\rm R}}}}
\def\IB{{\hbox{{\rm I}\kern-.2em\hbox{\rm B}}}}
\def\IN{{\hbox{{\rm I}\kern-.2em\hbox{\rm N}}}}
\def\IC{\,\,{\hbox{{\rm I}\kern-.59em\hbox{\bf C}}}}
\def\IZ{{\hbox{{\rm Z}\kern-.4em\hbox{\rm Z}}}}
\def\IP{{\hbox{{\rm I}\kern-.2em\hbox{\rm P}}}}
\def\IH{{\hbox{{\rm I}\kern-.4em\hbox{\rm H}}}}
\def\ID{{\hbox{{\rm I}\kern-.2em\hbox{\rm D}}}}
\def\II{{\hbox{\rm I}\kern-.2em\hbox{\rm I}}}

\noblackbox

\leftline{\epsfxsize1.0in\epsfbox{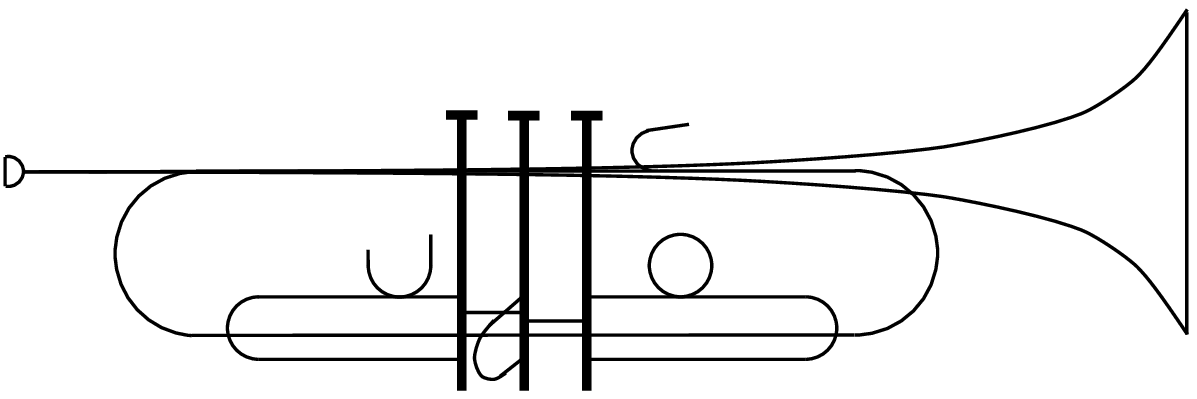}}
\vskip-1.0cm
\Title{\hbox{hep-th/9806115}}
{On Second--Quantized Open Superstring Theory}

\centerline{\bf Clifford V. Johnson$^\flat$}

\bigskip
\bigskip

\vbox{\baselineskip12pt\centerline{\hbox{\it 
Department of Physics}}
\centerline{\hbox{\it University of California}}
\centerline{\hbox{\it Santa Barbara CA 93106, U.S.A.}}}

\footnote{}{\sl email: $^\flat${\tt cvj@itp.ucsb.edu}$\,\,$
On leave from the Dept. of
Physics and Astronomy, University of Kentucky, Lexington KY~40502,
U.S.A.}

\vskip1.0cm
\centerline{\bf Abstract}
\vskip0.7cm
\vbox{\narrower\baselineskip=12pt\noindent
The $SO(32)$ theory, in the limit where it is an open superstring
theory, is completely specified in the light--cone gauge as a
second--quantized string theory in terms of a ``matrix string''
model. The theory is defined by the neighbourhood of a 1+1 dimensional
fixed point theory, characterized by an Abelian gauge theory with
type~IB Green--Schwarz form.  Non--orientability and $SO(32)$ gauge
symmetry arise naturally, and the theory effectively constructs an
orientifold projection of the (weakly coupled) matrix type~IIB theory
(also discussed herein). The fixed point theory is a conformal field
theory with boundary, defining the free string theory. Interactions
involving the interior of open and closed strings are governed by a
twist operator in the bulk, while string end--points are created and
destroyed by a boundary twist operator. }

\Date{14th June 1998}
\baselineskip13pt
\lref\super{C. V. Johnson, {\sl `Superstrings from Supergravity'},
 hep-th/9804200.}

\lref\green{M. B. Green and J. H. Schwarz, 
{\sl `Supersymmetrical String Theories'}, Phys. Lett. {\bf B109} (1982), 444.}

\lref\dbranes{J.~Dai, R.~G.~Leigh and J.~Polchinski, 
{\sl `New Connections Between String Theories'}, Mod.~Phys.~Lett.
{\bf A4} (1989) 2073\semi P.~Ho\u{r}ava, {\sl `Background Duality of
Open String Models'}, Phys. Lett. {\bf B231} (1989) 251\semi
R.~G.~Leigh, {\sl `Dirac--Born--Infeld Action from Dirichlet Sigma
Model'}, Mod.~Phys.~Lett. {\bf A4} (1989) 2767.}

\lref\banksreview{For a review, with references,
see {\sl `Matrix Theory'}, T. Banks, hep-th/9710231.}
\lref\sethisuss{S. Sethi and L. Susskind, {\sl `Rotational 
Invariance in the M(atrix) Formulation of Type IIB Theory'},
 Phys. Lett. {\bf B400} (1997) 265, hep-th/9702101.}

\lref\light{S. Mandelstam, {\sl `Interacting--String Picture of 
Dual--Resonance Models'}, Nucl. Phys. {\bf B64} (1973), 205; {\it
ibid.}, {\sl `Interacting--String Picture of the
Neveu--Schwarz--Ramond Model'}, Nucl. Phys. {\bf B69} (1974), 77\semi
E. Cremmer and J. L. Gervais, {\sl `Combining and Splitting
Relativistic Strings'}, Nucl. Phys. {\bf B76} (1974) 209; {\it ibid.}
{\sl `Infinite Component Field Theory of Interacting Relativistic
Strings and Dual Theory'}, Nucl. Phys. {\bf B90}, (1975) 410.  \semi
M. Kaku and K. Kikkawa, {\sl `Field Theory of Relativistic Strings,
I. Trees'}, Phys. Rev. {\bf D10} (1974) 1110; {\it ibid.}, {\sl `Field
Theory of Relativistic Strings, II. Loops and Pomerons'},
Phys. Rev. {\bf D10} (1974) 1823.  }
\lref\cone{M. B. Green and J. H. Schwarz, {\sl`Superstring Interactions'}, 
Nucl. Phys.  {\bf B218} (1983) 43; {\it ibid.,} {\sl`Superstring Field
Theory'}, Nucl. Phys. {\bf B243} (1984) 475.\semi M. B. Green,
J. H. Schwarz and L. Brink, {\sl`Superfield Theory of Type (II)
Superstrings'}, Nucl. Phys. {\bf B219} (1983) 437.}

\lref\aki{A. Hashimoto, {\sl `Dynamics of Dirichlet--Neumann
Open Strings on D-branes'}, Nucl. Phys. {\bf B496} (1997) 243,
hep-th/9608127.}

\lref\bits{R. Giles and C. B. Thorn, {\sl `A Lattice 
Approach To String Theory'}, Phys. Rev.  {\bf D16} (1977) 366\semi
C. B. Thorn, {\sl `Quark Confinement in the Infinite Momentum Frame'},
Phys. Rev. {\bf D19} (1979) 639; {\it ibid.,} {\sl `A Fock Space
Description of the $1/N_c$ Expansion of Quantum Chromodynamics'},
Phys. Rev. {\bf D20} (1979) 1435; {\it ibid.,} {\sl`Calculating the
Rest Tension for a Polymer of String Bits'}, Phys. Rev. {\bf D51}
(1995) 647, hep-th/9407169\semi I. Klebanov and L. Susskind,
{\sl`Contiuum Strings From Discrete Field Theories'}, Nucl. Phys. {\bf
B309} (1988) 175.}

\lref\heteroticcosets{C. V. Johnson, {\sl `Heterotic Coset Models'},
 Mod. Phys. Lett. {\bf A10} (1995) 549, hep-th/9409062\semi {\it
ibid.,} {\sl `Exact Models of Extremal Dyonic 4D Black Hole Solutions
of Heterotic String Theory'}, Phys. Rev.  {\bf D50} (1994) 4032,
hep-th/9403192.}

\lref\town{P. Townsend, {\sl `The eleven-dimensional 
supermembrane revisited'},
  Phys. Lett. {\bf B350} (1995) 184, hep-th/9501068.}

\lref\goed{E. Witten, {\sl `String Theory Dynamics in Various Dimensions'}, 
Nucl. Phys. {\bf B443} (1995) hep-th/9503124.}

\lref\morten{M. Krogh, {\sl `A Matrix Model for Heterotic
 $Spin(32)/Z_2$ and Type I String Theory'}, hep-th/9801034.}

\lref\bfss{T. Banks, W. Fischler, S. Shenker and L. Susskind, {\sl 
 `M--Theory As A 
Matrix Model: A Conjecture'}, Phys. Rev. {\bf D55}
 (1997) 5112, hep-th/9610043.}

\lref\dvv{R. Dijkgraaf, E. Verlinde, H. Verlinde, 
{\sl `Matrix String Theory'}, Nucl. Phys. {\bf B500} (1997) 43,
hep-th/9703030.}

\lref\edbound{E. Witten, {\sl `Bound States of Strings and $p$--Branes'}, 
Nucl. Phys. {\bf B460} (1996) 335, hep-th/9510135.}

\lref\gojoe{J. Polchinski, {\sl `Dirichlet Branes and Ramond--Ramond Charges
 in String Theory'}, Phys. Rev. Lett. {\bf 75} (1995) hep-th/9510017.}

\lref\eva{S. Kachru and E. Silverstein, {\sl `On Gauge Bosons in
 the Matrix Model Approach to M~Theory'}, Phys. Lett. {\bf B396} (1997)
70, hep-th/9612162.}
\lref\rey{N. Kim  and S-J. Rey, {\sl `M(atrix) Theory on an Orbifold and 
Twisted Membrane'}, Nucl. Phys. {\bf B504} (1997) 189,
hep-th/9701139\semi S-J. Rey, {\sl `Heterotic M(atrix) Strings and
Their Interactions'}, Nucl. Phys. {\bf B502} 170,1997,
hep-th/9704158.}
\lref\banksmotl{T. Banks and L. Motl, 
{\sl `Heterotic Strings from Matrices'}, JHEP 12 (1997) 004, hep-th/9703218.}
\lref\matrixheterotic{L. Motl, {\sl `Quaternions and M(atrix) Theory
in Spaces with Boundaries'}, hep-th/9612198\semi T. Banks, N. Seiberg,
E. Silverstein, {\sl `Zero and One-dimensional Probes with N=8
Supersymmetry'}, Phys. Lett. {\bf B401} (1997) 30, hep-th/9703052\semi
D. Lowe, {\sl `Bound States of Type I' D-particles
and Enhanced Gauge Symmetry'}, Nucl. Phys. {\bf B501} (1997) 134,
hep-th/9702006; {\it ibid.}, {\sl `Heterotic Matrix String Theory'},
Phys. Lett. {\bf B403} (1997) 243, hep-th/9704041.}
\lref\petr{P. Horava, {\sl `Matrix Theory and
Heterotic Strings on Tori'}, Nucl.  Phys. {\bf B505} 84 (1997),
hep-th/9705055.}
\lref\petrorientifold{P. Horava, {\sl`Strings on World--Sheet Orbifolds'},
 Nucl. Phys. {\bf B327} (1989) 461.}
\lref\kabat{D. Kabat and
S-J. Rey, {\sl `Wilson Lines and T-Duality in Heterotic M(atrix)
Theory'}, Nucl. Phys. {\bf B508} 535, (1997) hep-th/9707099.}
\lref\matrixheteroticii{S. Govindarajan, {\sl `Heterotic
M(atrix) theory at generic points in Narain moduli space'},
hep-th/9707164.}
\lref\motl{L. Motl, {\sl `Proposals on Non--Perturbative Superstring
 Interactions'}, hep-th/9701025.}
\lref\banks{T. Banks and N. Seiberg, {\sl `Strings from Matrices'},
 Nucl. Phys. {\bf B497} 41 (1997), hep-th/9702187.}
\lref\danielsson{U. H. Danielsson, G. Ferretti, {\sl `The Heterotic
 Life of the D-particle'}, Int. J. Mod. Phys. {\bf A12} 
(1997) 4581, hep-th/9610082.}

\lref\anatomy{C. V. Johnson, {\sl`Anatomy of a Duality'}, 
Nucl. Phys. {\bf B521} (1998) 71, hep-th/9711082.}

\newsec{Opening Remarks}
\subsec{Motivations}

In the earliest searches for a better understanding of the
non--perturbative structure of string theory, there was considerable
attention given to finding a second--quantized description of the
theory\refs{\light,\cone}. Considering the logic of the development of
quantum field theory, this was a plausible next step.

As it turned out, the powerful tools which may be exploited in the
presence of extended supersymmetry led to the discovery that
``strong/weak coupling duality'' was a compelling hint about the
non--perturbative nature of string theory\refs{\town,\goed}. This
eventually led to the simplest possible statement about the five ten
dimensional string theories and the meaning of their duality
relationships: The string theories are all ``weak coupling'' limits of
a single parent theory, ``M--Theory'', and the duality operations
connecting them are consequences of geometrical relationships between
the limits.

The search for a full definition of M--theory continues. A partial one
is given in terms of ``Matrix Theory''\bfss, which has been shown to have
many of the exciting and peculiar properties required by a theory
which must play such a distinguished role\banksreview.

Among these properties are (of course) the recovery of the five weakly
coupled superstring theories in the appropriate limits.  These
theories are {\sl all}\foot{As argued in ref.\super, by organizing and
extending results of
refs.\refs{\motl,\sethisuss,
\banks,\banksmotl,\dvv,\eva,\danielsson,\matrixheterotic}}\
described in terms of the neighbourhood of a~1+1~dimensional orbifold
conformal field theory. The strings are free in the conformal field
theory limit, and interactions are turned on by turning on a twist
operator in the theory. The operator is irrelevant, and therefore
deforms the theory away from the fixed point.

The conformal field theories are characterized in Lagrangian form by a
matrix--valued version of the appropriate Green--Schwarz action for the
string. The matrices are diagonal, and the permutations of diagonal
elements is a gauge symmetry of the model. The orbifold target space
of the conformal field theory is the gauge invariant vacuum moduli space
of this theory, and long strings arise in the twisted sectors of the
orbifold.

Away from the conformal field theory limit, where interactions have
been turned on, the theories cease to have such similar descriptions,
as required by duality.  Best understood are the type
IIA\refs{\motl,\banks,\dvv}\ and $E_8{\times}E_8$
heterotic\refs{\danielsson,\eva,\matrixheterotic,\rey}\ string
theories, because the interaction operator takes one back along the IR
flow to a 1+1 dimensional non--Abelian version of the (relevant)
models described in the previous paragraph (the matrices are no longer
diagonal).  Furthermore, for extreme values of the string coupling,
the theory returns to a 0+1 dimensional matrix quantum mechanics.

By contrast, the type~IIB and the two $SO(32)$ theories do not become
1+1 dimensional non--Abelian matrix Green--Schwarz models for
arbitrary values of the string coupling, but instead become (at
intermediate values) 2+1 dimensional fixed point
theories\refs{\sethisuss,\banks,\super,\petr,\kabat}, which are so far
not very well studied. At strong coupling, they become 1+1 dimensional
again, and ultimately give the appropriate weakly coupled dual string
theory\super.

A pleasant and perhaps ironic feature of the neighbourhood of weak
coupling in all five cases is that the description is intrinsically
a second--quantized theory of strings in the light--cone gauge. In
this way, we see just how the early studies of light--cone string
field theory fit into the modern scheme of things. From the previous
two paragraphs, it is clear from this perspective just how the physics
encoded in weak coupling (light--cone) string field theory is
completed into non--perturbative information consistent with the
duality results.

(For a few more comments on the role of string field theory in a
modern context, see the closing remarks (section~5).)

The purpose of this paper is to make explicit the matrix realization
of the second--quantized description of the $SO(32)$ ``type~IB''
string theory. This stringy description is a novel 1+1 dimensional
theory which was discovered in ref.\super\ by taking the appropriate
limits of the matrix description of M--theory on a
cylinder\foot{Aspects of the description of the free theory were
actually found earlier in ref.\banksmotl\ in the context of a
description of the $E_8{\times}E_8$ heterotic and type~IA strings. I
thank L. Motl for pointing this out to me.}. It is necessarily
different from the description of the type~II and heterotic theories,
and therefore perhaps deserves some pedagogy.

\subsec{Summary of Results}

$\bullet$ The free strings are described by a conformal field theory
with boundary. This conformal field theory is characterized by an
$N{\times}N$ matrix Green--Schwarz action for the type~IB string. The
action is simply the matrix Green--Schwarz action for the type~IIB
string, with reflecting boundary conditions inserted.

$\bullet$ The second--quantized Fock space contains both open and
closed strings with momentum in the light--cone direction proportional
to their length. These strings are built out of winding type~IA
strings glued together in the twisted sectors of the orbifold to make
long strings which survive the large $N$ limit, defining the
infinite momentum light--cone frame.

$\bullet$ The long strings are unoriented. By examining the strings
which arise in the twisted sectors, it is clear that the strings are
non--oriented. The construction ensures that each sequence of
constituent type~IA string bits of a given orientation is accompanied
by the same structure with the opposite orientation.

$\bullet$ The open strings carry an $SO(32)$ gauge symmetry. This
follows from the non--perturbative origins of the model\super: The
model is built on a type~IA fundamental string background, which has
eight D8--branes and an O8--plane at each end of the interval.  The
required $SO(32)$ quantum numbers are explicitly assembled out of
$SO(16){\times}SO(16)$ quantum numbers: For example, the adjoint of
$SO(32)$ is filled out by open strings of even length, which transform
as ${\bf(1,120)}{\oplus}{\bf(120,1)}$, together with open strings of odd
length, which fill out $\bf(16,16)$.

$\bullet$ Interactions involving the interior of open and closed
strings are governed by a vertex operator analogous to that presented
in ref.\dvv. The coefficient of the operator is proportional to the
closed string coupling $g^{\phantom{2}}_{\rm IB}$.

$\bullet$ Interactions involving only  string end--points  are
also governed by a vertex operator, but this operator lives on the
boundary of the conformal field theory. The operator also decorates
the ends of strings with the Chan--Paton factors which live on the
boundary.

The strength of this open--open operator is proportional
to the square root of the closed string coupling, as is appropriate
for a Yang--Mills coupling: $g^{\phantom{2}}_{\rm YM}{\sim}g_{\rm
IB}^{1/2}$. This is of course the string field theory statement that
type~IB strings end on D9--branes.

\newsec{Second--Quantized  Type~IIB Strings}
We begin with an explicit description of the matrix theory of the
weakly coupled type~IIB string, emphasizing its similarities to and
differences from its type~IIA counterpart.

Consider the following action:
\eqn\greenschwarzi{S={1\over 2\pi}
\int d^2\sigma \,{\rm Tr}\biggl((\partial_a 
X^i)^2-i\Theta^T\gamma^a \partial_a\Theta\biggr).}  The $X^i$ are
eight scalar fields, ($i{=}1,{\ldots},8$), and $\Theta$ contains two
Majorana--Weyl fermionic fields $\theta_L^\alpha$ and
$\theta_R^\alpha$ which respectively transform in the ${\bf 8}_v$,
${\bf 8}_s$ and ${\bf 8}_s$ (vector and spinor; the conjugate spinor
is denoted~${\bf 8}_c$) representations of the $SO(8)$ R--symmetry of
the model. They are all diagonal $N{\times}N$ hermitian matrices. The
world--volume coordinates are
$\tau{\equiv}\sigma^0,\sigma{\equiv}\sigma^1$, with
$0\leq\sigma\leq2\pi$.


The action \greenschwarzi\ is invariant under the 16 supersymmetries:
\eqn\supersymm{\eqalign{\delta X^i&= 
{2\over\sqrt{2p^+}}{\bar\epsilon}\Gamma^i\Theta\cr
\delta \Theta&={i\over\sqrt{2p^+}}\Gamma^-\Gamma^i
\gamma^a\partial_aX^i\epsilon}}
where $\epsilon$ contains $\epsilon_L^\alpha$ and
$\epsilon_R^\alpha$. The $\Gamma$ and $\gamma$ are ten dimensional and
two dimensional gamma matrices, respectively.

The constant $p^+$ has the interpretation of light--cone (IMF)
momentum once we recall the identification of the remaining spacetime
directions $X^\pm{=}(X^0\pm X^9)/\sqrt{2}$ with the world--sheet time:
$X^+(\sigma,\tau){=}x^+{+}p^+\tau$.

This model is thus far simply $N$ copies of the type~IIB
Green--Schwarz action\green. (We may also think of it as $N$ copies of
the type~IIA string in a {\it static} gauge, aligned along a space--like
direction which we will call $X^9$, with radius $R_9$.)

There are two supercharges $Q_{L(R)}^{\dot\alpha}$:
\eqn\charges{Q_{L(R)}^{\dot\alpha}=
{i\over\pi}{1\over\sqrt{2p^+}}\int_0^{2\pi}d\sigma\, {\rm
Tr}\left[\Gamma^i_{\alpha{\dot\alpha}}(\gamma^a\partial_aX^i(\sigma))
\gamma^0\theta^\alpha_{L(R)}(\sigma)
\right]}
representing the 16 non--trivial supersymmetries of the model. The other
16 are trivially realized  as: 
\eqn\trivial{\delta X^i{=}0,\quad\delta\Theta=\sqrt{2p^+}\eta.}

This model has a discrete gauge symmetry $S_N$, corresponding to the
freedom to permute the $N$ eigenvalues of the matrices. The moduli
space of this theory is characterized by the manifold of gauge
inequivalent values of the bosonic fields as simply:
\eqn\target{{\cal M}_{\rm IIB}\equiv{(\IR^8)^N\over S_N}.}
This space is of course the same as that for the type~IIA string, but
here, the fermions transform appropriately for the type~IIB string.

We present this as the target space of a free orbifold conformal field
theory. The untwisted sectors of this orbifold correspond to $N$
independent (light--cone gauge) type~IIB strings with a single unit of
momentum in $X^+$. Explicitly, this represents $N$ (static gauge)
type~IIA strings with 1 unit of winding around $X^9$. The twisted
sectors of the orbifold are classified in terms of the conjugacy
classes of $S_N$, which are labeled by the number $n$ of eigenvalues
which are permuted by a given representation. The twisted sector $n$
defines a type~IIA string of length $n$, {\it i.e.,} it is wound $n$ times
around the circle. This in turn defines a type~IIB string with $n$
units of momentum in the $X^+$ direction.

Explicitly, a configuration representing a string of length $n$ is
described by (sub--) matrices of the form:
\eqn\matrixformi{X^i(\sigma)=\pmatrix{x^i_1(\sigma)&0&0&\ldots&0\cr
0&x^i_2(\sigma)&0&\ldots&0\cr
0&0&x^i_3(\sigma)&\ldots&0\cr\vdots&\vdots&\vdots&\ddots&\vdots\cr
0&0&0&0&x^i_n(\sigma)},}
with
\eqn\conditioni{X^i(\sigma+2\pi)=S_nX^i(\sigma)S_n^{-1},}
where
\eqn\permute{S_n=\pmatrix{0&0&0&\ldots&1\cr1&0&0&\ldots&0\cr
0&1&0&\ldots&0\cr\vdots&\vdots&\ddots&\ddots&\vdots\cr
0&0&0&1&0}}
is an $n{\times}n$ (sub--) matrix which permutes the eigenvalues.
In other words:
\eqn\condii{X^i(\sigma+2n\pi)=X^i(\sigma).}
A diagram of a closed string of length~3 is depicted in fig.1:

\midinsert{
\vskip0.1cm
\centerline{\epsfxsize4.5in\epsfbox{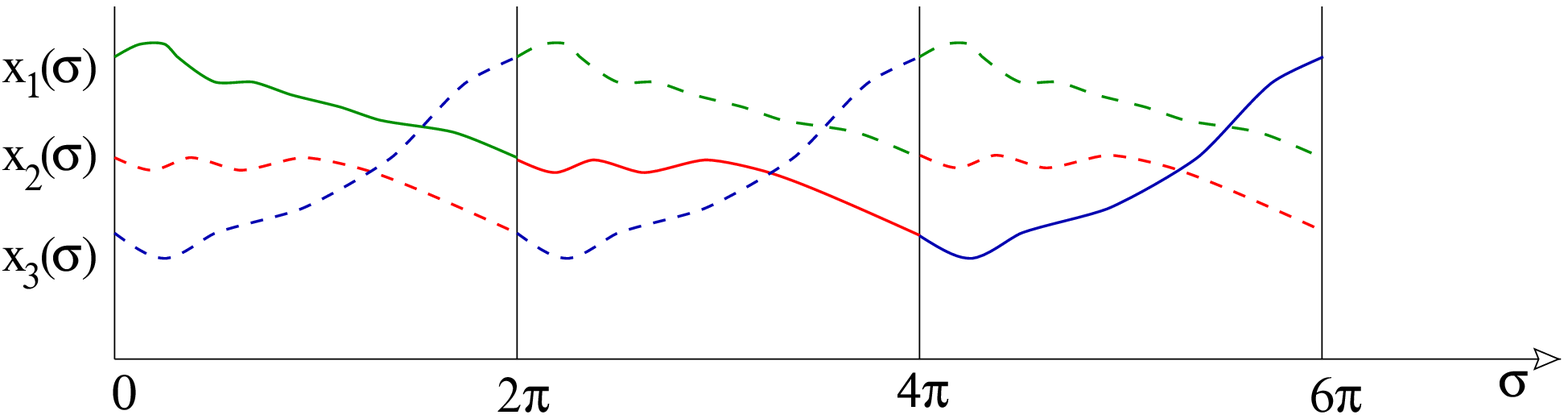}}
\vskip0.1cm
\centerline{\sl Figure 1.}}
\endinsert

The eigenvalues $x^i_j(\sigma)$, (which sort of play the role of
``string bits''\bits) are otherwise {\sl arbitrary} functions of
$\sigma$ and therefore the complete Fock space of the two dimensional
field theory contains type~IIB strings of arbitrary shape in $\IR^8$,
in the various discrete momentum sectors. This is why this is a
second--quantized description of the strings.

Ultimately, we want the large $N$ limit. We make long strings of
length $n$ by keeping the ratio $n/N$ fixed as we send
$N{\to}\infty$. This defines for us the light--cone gauge (IMF)
type~IIB string, where $n/N$ is the finite fraction of the total $p^+$ which
the long string possesses. As shown in ref.\dvv, the correct string
tension and level matching for the long strings emerges correctly. We
will not repeat this here.

If this was the type~IIA string, we could think of the Lagrangian
above as the strong coupling limit of a non--Abelian matrix
Green--Schwarz theory (such as would be derived from the world--volume
of $N$ coincident D1--branes). This is the 1+1 dimensional gauge
theory from which the free type~IIA matrix theory
flows\refs{\dvv,\edbound}:
\eqn\greenschwarzii{S={1\over 2\pi}
\int d^2\sigma {\rm Tr}\biggl((D_\mu X^i)^2-i\Theta^T\gamma^\mu D_\mu\Theta
+{g}^2_{\rm IIA}F^2_{\mu\nu} -{1\over{g}^2_{\rm
IIA}}[X^i,X^j]^2+{1\over {g}^{\phantom{.}}_{\rm
IIA}}\Theta^T\Gamma_i[X^i,\Theta]\biggr).}  The dimensionless coupling
${g}^{\phantom{.}}_{\rm IIA}$ is the matrix type ~IIA string coupling
inversely proportional to the 1+1 dimensional Yang--Mills coupling.

In the present case, the situation is more complicated.  The free
type~IIB matrix Green--Schwarz Lagrangian does {\it not} flow from
such an interacting 1+1 gauge theory, but arises as a reduction from a
2+1 dimensional fixed point theory. (This is consistent with the fact
that type~IIA does not contain D1--branes.)

We do not have a gauge theory definition of the theory analogous to
eqn.\greenschwarzii\ even slightly away from the 2D fixed point\foot{I
thank P. Pouliot for a discussion on this point.}\ but for weak string
coupling, where we are close to the free conformal field theory limit,
we can describe non--zero string coupling in terms of giving an
expectation value to a certain operator in the conformal field theory.

The operator is very similar to the one presented in ref.\dvv, for the
type~IIA case. There, the joining of the closed strings is described
by the exchange of eigenvalues between configurations representing two
different strings (see fig.~6). (The reverse procedure will give the
splitting of a closed string into two closed strings.) This can be
written as a $\IZ_2$ twist operator in the field theory. The string
interaction vertex is built out of twist fields in the conformal field
theory.

A similar construction will give the interaction vertex for the
type~IIB case, the only difference being that the left-- and
right--moving parts of it (and structures which descend from them)
will have identical individual $SO(8)$ transformations. We will return
to this operator later in section~4.1, as it will appear in the
construction of the interactions of unoriented closed and open
strings.

\newsec{Second--Quantized  Type~IB Strings}
The construction of ref.\super\ showed that the matrix string
description of the type~IB string is given in terms of winding type~IA
strings. 

The type~IA model is characterized by the orientifold group
$\{1,\Omega {\cal R}_9\}$, where ${\cal R}_9$ is the reflection
$X^9{\to}{-}X^9$ and $\Omega$ exchanges left-- and right--moving
fields. In other words, given that $\sigma$ is identified with $X^9$
(static gauge) we must study a model similar to the above, but with
the $\sigma$ circle $S^1$ replaced by the orbifold~$S^1/\IZ_2$.

The fundamental domain of $\sigma$ is now $[0,\pi]$ instead of
$[0,2\pi]$, and there are fixed points of $\Omega{\cal R}_9$ at
$\sigma{=}0$ and $\sigma{=}\pi$. There is an O8--plane and 8
D8--branes at each end of the interval. Essentially the Lagrangian we
seek will be associated with a macroscopic type~IA string stretched
along the interval.

Eventually, just as we worked on the covering space of the circle to
make the long type~IIB strings, we will work on the covering space of
the orbifold which has a description as follows: Break the infinite
line up into segments of length $2\pi$. Half of that segment is a copy
of the fundamental domain described above, while the other half is
also a copy, but mirror reflected. This unit then repeats
periodically. String bits will live on this covering space, subject to
the reflection conditions (see figs.~2 and~5).

We define the Lagrangian of the model by imposing  conditions
upon the Lagrangian \greenschwarzi\ consistent with the above description of
the interval. 

We (initially) impose that the fields  satisfy:
\eqn\conditionii{\eqalign{X^i(2\pi-\sigma)&= X^i(\sigma)\cr
\theta^\alpha_{L(R)}(2\pi-\sigma)&=\theta^\alpha_{R(L)}(\sigma)
.}} 

\subsec{Short Strings}
Condition \conditionii\ defines two types of short string.

$(i)$ One type of string is an open string stretched along the interval
$(0,\pi)$, with boundary conditions on a single eigenvalue function
(and its superpartner):
\eqn\boundary{\eqalign{{\partial x^i(0,\tau)/\partial\sigma}=0\quad{\rm and}&
\quad{\partial x^i(\pi,\tau)/\partial\sigma}=0\cr
\theta^\alpha_{L}(0,\tau)=\theta_{R}^\alpha(0,\tau)\quad{\rm and}&
\quad\theta^\alpha_{L}(\pi,\tau)=\theta_{R}^\alpha(\pi,\tau).}}
This immediately tells us that the strings which survive the
projection of the last section are open strings (of length~1), with
endpoints on the ends of the interval. This is as we might expect, as
we have simply imposed type~IB boundary conditions on the
Green--Schwarz action.

$(ii)$ Another type of string can be seen by working on the doubled cell
$0<\sigma\leq2\pi$. Now we see that we can define a closed string of
length 1 by covering the type~IA interval twice, using one eigenvalue:
\eqn\short{\eqalign{x^i(2\pi-\sigma)=&x^i(\sigma)\cr
\theta_{L(R)}^i(2\pi-\sigma)=&\theta_{R(L)}^i(\sigma).}}

\midinsert{
\vskip0.1cm
\centerline{\epsfxsize2.0in\epsfbox{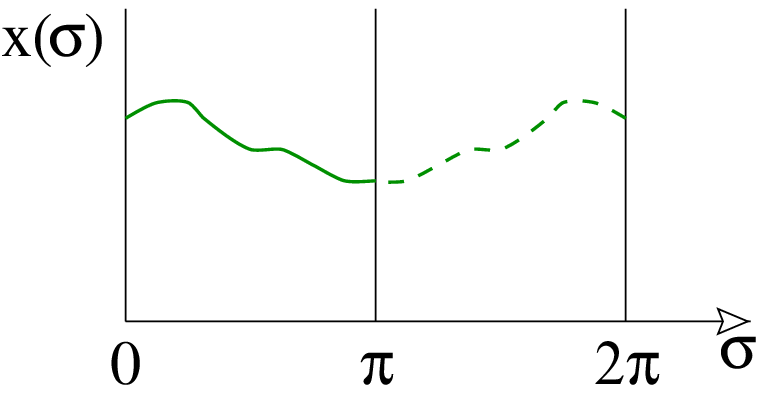}}
\vskip0.1cm
\centerline{\sl Figure 2.}}
\endinsert

We may think of the illustration in fig.~2 in two ways, therefore. In
both ways the two halves of the diagram may be thought of as two parts
of a string. The first way depicts the left and right moving parts of
a string, which are identical, and are put together to make the open
string.  The second way of thinking describes both halves as sections
of a closed string. Because they are identical in shape, except for a
reflection, the resulting closed string is unorientable.

Notice that these strings have only ${\cal N}{=}1$ ten dimensional
spacetime supersymmetry because we have reduced the number of
supercharges by one half by identifying the $\theta_L$ and $\theta_R$
as we cross $\sigma{=}\pi$. The remaining supercharge is:
\eqn\charges{Q^{\dot\alpha}={1\over2}(Q^{\dot\alpha}_L+Q^{\dot\alpha}_R).}

So we see that two types of string descend from our oriented type~IIB
string: the unoriented open and closed strings. They both come from
identifying left and right, the former with two fixed points on the
strings ---giving end--points--- and the latter without.

One can think of all open and unoriented string diagrams in this
way. For example, at tree level the sphere ${\rm S}^2$ can be
projected to give the disc ${\rm D}^2$ and the real projective
plane~$\IR {\rm P}^2$ (See fig.~3).

\midinsert{
\vskip0.1cm
\centerline{\epsfxsize1.5in\epsfbox{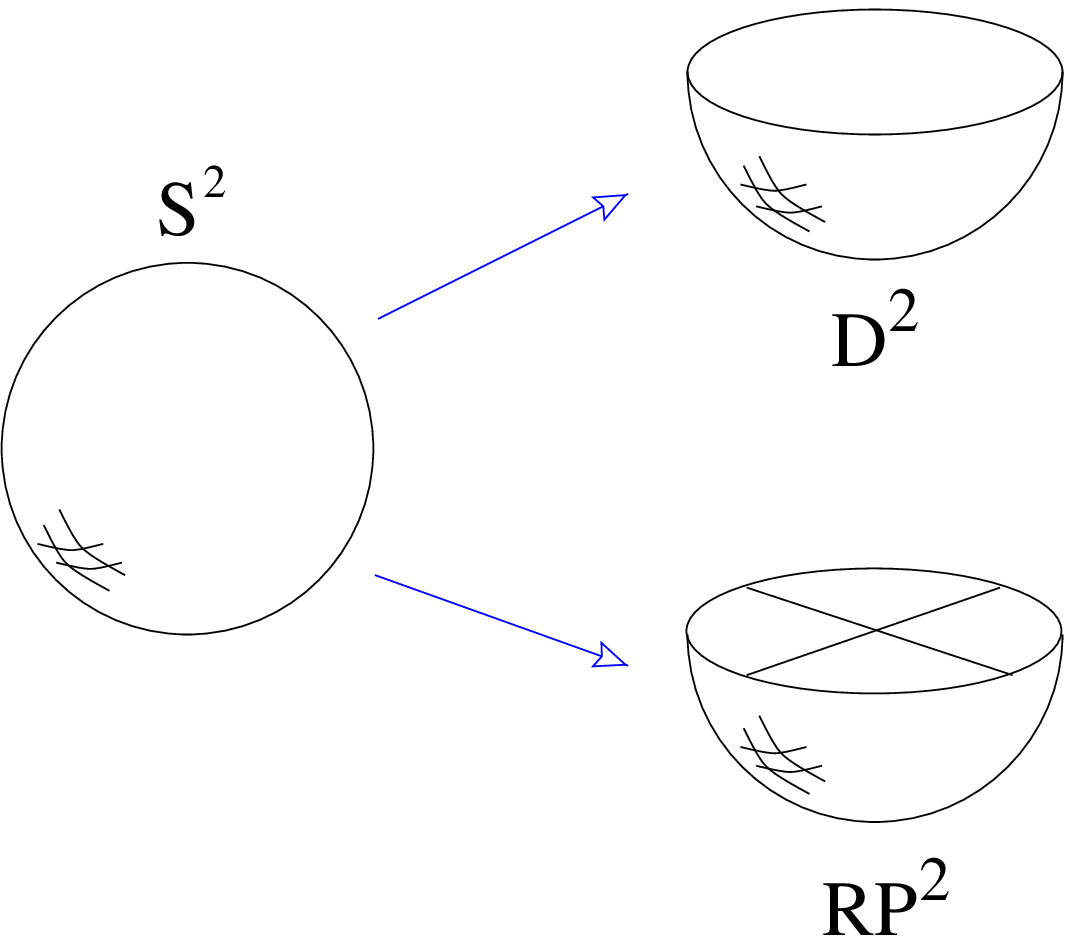}}
\vskip0.1cm
\centerline{\sl Figure 3.}}
\endinsert

Meanwhile, at one loop, the torus ${\rm T}^2$
projects to give the cylinder ${\rm C}^2$, the M\"obius strip ${\rm
MS}$ and the Klien bottle ${\rm KB}$ (see fig.~4):

\midinsert{
\vskip0.1cm
\centerline{\epsfxsize2.0in\epsfbox{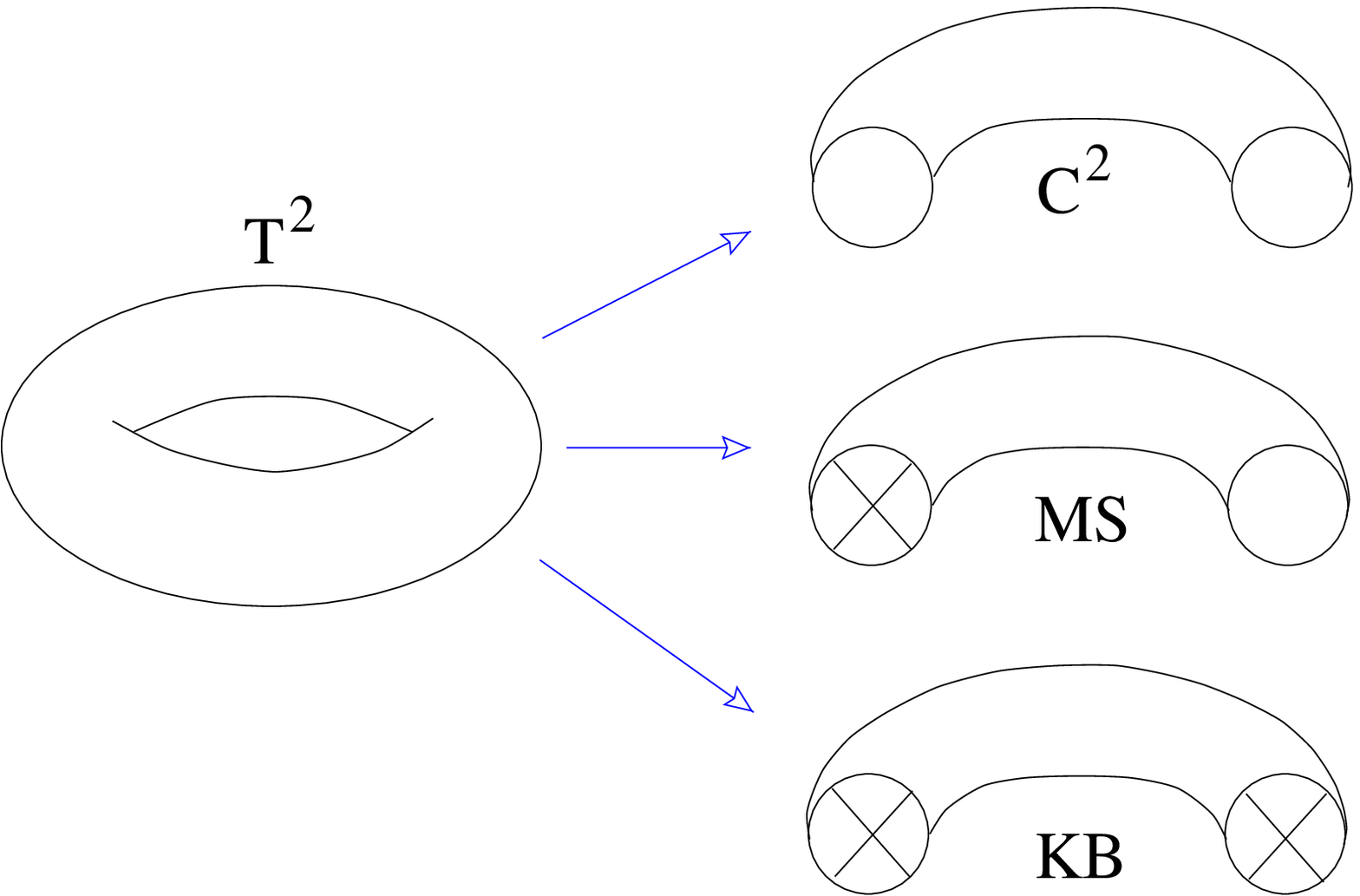}}
\vskip0.1cm
\centerline{\sl Figure 4.}}
\endinsert

So altogether we have $N$ independent static gauge type~IA strings of
length 1, or $N$ light--cone type~IB strings each with a single unit of
momentum in the $X^+$ direction.

\subsec{Long Strings}
Of course, the magic of this construction is that we can exploit the
discrete gauge symmetry $S_N$ to construct long strings from the
twisted sectors of the orbifold. This translates into considering
twisted versions of the conditions \conditionii.

To make a long string of length $n$, we again exploit a permutation of
$n$ eigenvalues, $S_n$, every time we translate by $2\pi$. In
addition, however, we must obey the requirement of $\IZ_2$ reflection
symmetry \conditionii\ within each closed string unit cell.

In general, the matrix $X^i$ will be made of $n$ different eigenvalue
functions $x_1^i,\ldots,x^i_n$, which each decorate the covering space
in a way which is $2\pi$ periodic and $\IZ_2$ symmetric about every
multiple of $\pi$. The $S_n$ permutation makes the matrix
$X^i(\sigma)$ a $2n\pi$ periodic matrix, defining either a long closed
string of length $n$, or a long open string of length $n$ and its
mirror.  For example, a closed (or open) string of length $3$ is shown
in fig.~5:

\midinsert{
\vskip0.1cm
\centerline{\epsfxsize4.5in\epsfbox{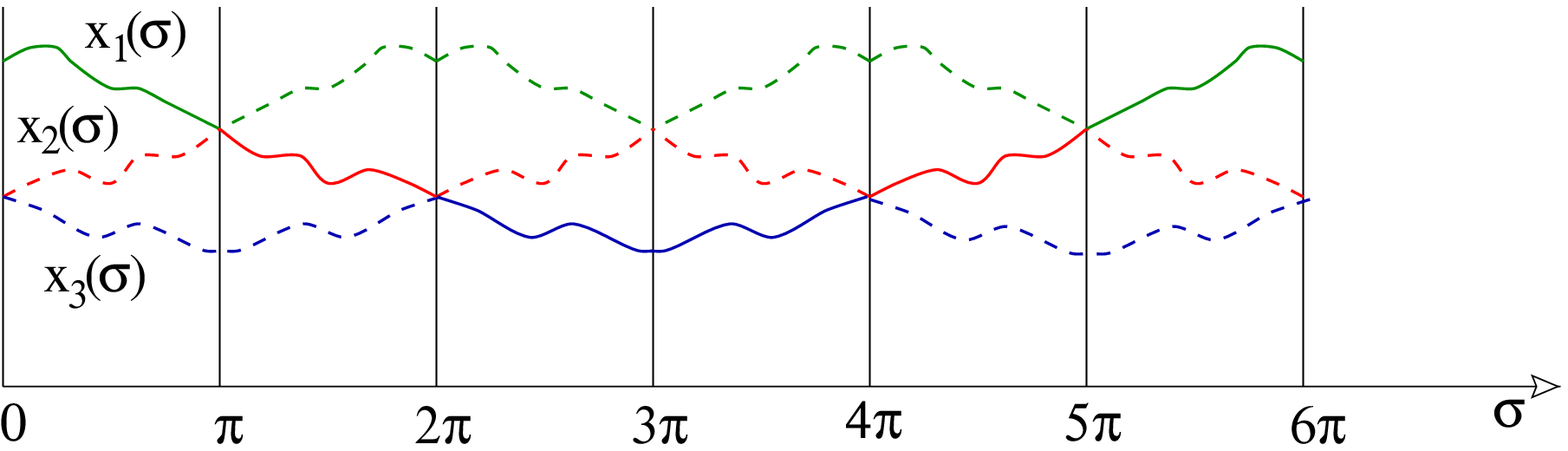}}
\vskip0.1cm
\centerline{\sl Figure 5.}}
\endinsert

Notice that by construction any sequence of string bits of a given
shape is accompanied by an identical sequence of the mirror reflected
shape somewhere along the string in such a way that the resulting
closed strings which are made by this procedure are guaranteed to be
unorientable.

Another way to characterize the twisted sectors we have found is as
follows: We must recall that there is a left--right exchange symmetry
$\Omega_n$ of {\sl all} long closed strings which we can make in the
type~IIB matrix model. We may think of it as the $\Omega$ which
already exists for each short type~IIB string ({\it i.e.,} it acts on
one eigenvalue) combined with a permutation of $n$ eigenvalues $S_n$
in such a way as to define $\Omega_n$, the exchange of left and
right--moving parts of the length $n$ string which is periodic over
$0<\sigma\leq2n\pi$.

In this way, we see that long unoriented closed and open strings arise
as twisted sectors of the orbifold theory. Again, asking that we keep
finite fractions $n/N$ fixed in the large $N$ limit, we construct the
complete Fock space of the free type~IB strings in light--cone gauge.

\subsec{Spacetime $SO(32)$ Gauge Symmetry}

One last property of our second quantized Fock space is that the open
string end--points carry $SO(32)$ gauge symmetry. This is easy to
show.

First of all, the limits which led\super\ to the construction of this
model from a non--perturbative framework very specifically took us to
a type~IA background, which has 8 D8--branes and~1~O8--plane at each
end of the interval. We have therefore a manifest
$SO(16){\times}SO(16)$ gauge symmetry as a background, but the Fock
space will assemble $SO(32)$ out of this as follows. There are two
classes of long open string:

$(i)$ Those which have endpoints only at one
or other end of the interval. These are strings with $n$ even. This
class will give states in the ${\bf(120,1)}{\oplus}{\bf(1,120)}$, the
adjoint of  $SO(16){\times}SO(16)$.

$(ii)$ Those which have one end on each
interval. These are strings with $n$ odd. These will give the
bi--fundamental $\bf(16,16)$ of $SO(16){\times}SO(16)$.

The presence of all of these long strings in the large $N$ Fock space
means that we can assemble the 496 dimensional adjoint of $SO(32)$,
showing that our light--cone open string spectrum indeed has the
correct gauge symmetry.

It is amusing to see $SO(32)$ arise in this way, as it is the precise
analogue of what took place in realizing $E_8{\times}E_8$ gauge
symmetry for the matrix heterotic string. 

In the 0+1 dimensional picture, the discussion was phrased\eva\ in
terms of bound states of D0--branes. Bound states of even numbers of
D0--branes in the neighbourhood of either (1~O8--plane+8~D8--brane)
unit gave the adjoint (${\bf(120,1){\oplus}(1,120)}$) of
$SO(16){\times}SO(16)$ while bound states of odd numbers of D0--branes
gave the spinor (${\bf(128,1){\oplus}(1,128)}$) filling out the adjoint
(${\bf248{\oplus}248}$) of $E_8{\times}E_8$.

In the 1+1 dimensional picture, this was translated\banksmotl\ into
periodic (P) and anti--periodic (A) boundary conditions on the
heterotic fermions (arising from D1--D9 strings in the presence of an
$SO(32){\to}SO(16){\times}SO(16)$ Wilson line). In constructing long
heterotic strings, the standard heterotic GSO projection assembled
$E_8{\times}E_8$ out of the adjoints (AA) (even length strings) and
the spinors (AP) (odd length strings).

\newsec{Interacting Second--Quantized Type~IB Strings}

\subsec{String Bulk Interactions}
As previously discussed in the case of the type~IIA and
$E_8{\times}E_8$ heterotic strings\refs{\dvv,\rey}, the joining of
closed strings is controlled by the exchange of eigenvalues between
two configurations representing separate closed strings. This is a
simple $\IZ_2$ operation on the coordintates $x_-{\equiv}x_1{-}x_2$
and $\theta_-{\equiv}\theta_1-\theta_2$:
$(x_-,\theta_-){\to}(-x_-,-\theta_-)$, which can occur for any value
of $\sigma$, {\it i.e.} anywhere on the closed string (see fig.~6).
(Obviously, this also controls the reverse, {\sl splitting}, process.)

\midinsert{
\vskip0.1cm
\centerline{\epsfxsize3.5in\epsfbox{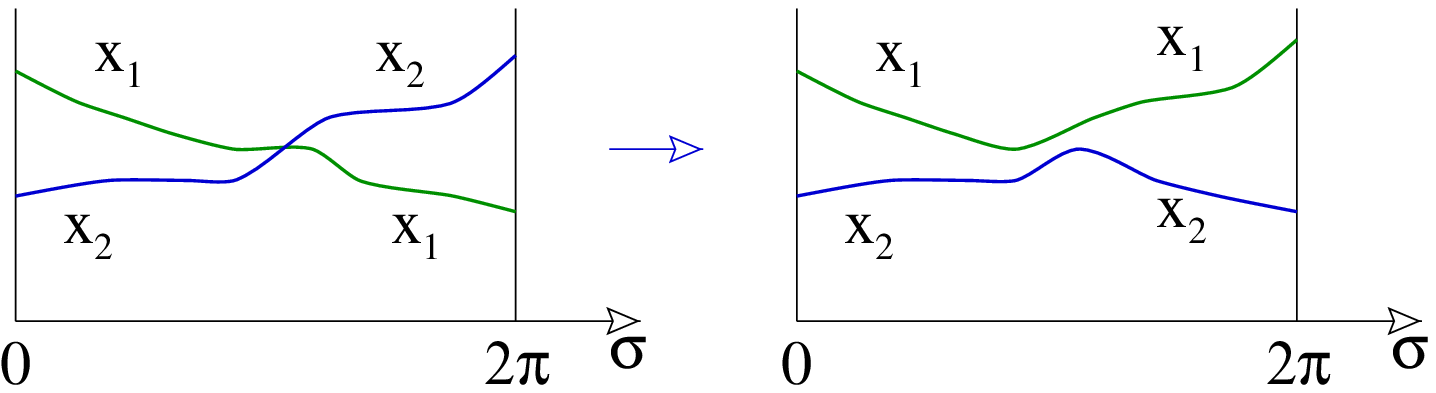}}
\vskip0.1cm
\centerline{\sl Figure 6.}}
\endinsert

In the conformal field theory, the interaction is introduced by adding
an operator which is constructed out the standard $\IZ_2$ conformal
field theory twist fields.

Let us recall the discussion of ref.\dvv.
The following operator product:
\eqn\opei{x^i_-(z)\cdot\sigma(w)\sim (z-w)^{-{1\over2}} \tau^i(w)} 
defines the standard twist field $\sigma(z)$ and its conjugate partner
$\tau^i(z)$, which have conformal dimension
$8{\times}{1\over16}{=}{1\over2}$ and $1$, respectively. The field
$\tau^i(z)$ transforms in the vector ${\bf 8}_v$ of $SO(8)$. (We use
complex world sheet coordinate $z{=}\exp(\tau{+}i\sigma)$, defining
the complex plane. Our light--cone strings are matrices wound on
circles centered on the origin. Time runs radially.)

The operators products:
\eqn\opeii{\eqalign{&\theta_-^\alpha(z)\cdot\Sigma^i(w)\sim 
(z-w)^{-{1\over2}}\Gamma^i_{\alpha{\dot\alpha}}\Sigma^{\dot\alpha}(w)\cr
&\theta_-^\alpha(z)\cdot\Sigma^{\dot\alpha}(w)\sim
(z-w)^{-{1\over2}}\Gamma^i_{\alpha{\dot\alpha}}\Sigma^i(w) }} define
the spin fields $\Sigma^i$ and $\Sigma^{\dot\alpha}$, which transform
in the vector ${\bf 8}_v$ and the conjugate spinor~${\bf 8}_c$
respectively.

The interaction vertex is constructed out of the field
\eqn\field{{\cal O}(z)=[\tau^i\Sigma^i](z),}
which is manifestly $SO(8)$ invariant. 

For the type~IIA theory, the full operator is constructed by
performing the same construction on the right hand side, but with the
occurrences of fields in ${\bf 8}_s$ and ${\bf 8}_c$ $SO(8)$
representations exchanged.

In the case in hand, we wish to construct first the type~IIB operator,
and so the same $SO(8)$ spinor representations appear for the fermions
and spin fields on the right hand side as above. At the end of the
day, the actual vertex is constructed out of vectors ${\bf 8}_v$, and
so it is particularly simple to write the $SO(8)$ invariant
supersymmetric operator. For completeness, we must sum over all the
possible eigenvalues (labeled $I,J$) which the operator will permute
(as was done for type~IIA), to give the interaction:
\eqn\interaction{\lambda\int 
d^2z\sum_{I<J}
\left(\tau^i\Sigma^i\otimes{\bar\tau}^j{\bar\Sigma}^j\right)_{IJ}.}

As this is a wieght $\left({3\over2},{3\over2}\right)$ field, the
coupling $\lambda$ has conformal dimension $-1$, and is linear in the
matrix string coupling\dvv:\eqn\scale{\lambda\sim g^{\phantom{2}}_{\rm
IIB}\ell_s.}

As the interaction operator ${\cal O}(z,{\bar z}){=}{\cal
O}(z){\bar{\cal O}}({\bar z})$ (and its possible descendants) has
exactly the same structure on the left and right, we readily see that
we can construct a symmetrized version of it:
\eqn\proj{{\widetilde{\cal O}}(z,{\bar z})
={1\over2}\left({\cal O}(z){\bar{\cal O}}({\bar z})+{\cal O}({\bar
z}){\bar{\cal O}}(z)\right)} to give a bulk operator which controls
the splitting and joining of unoriented strings somewhere along their
length (see fig~7(b)). This operator naturally lives on $\IR{\rm
P}^2$, and the complex $z-$plane we are working on is a double cover
of it.

This operator ${\widetilde{\cal O}}$ also has dimension
$\left({3\over2},{3\over2}\right)$, and so its coupling $\lambda_1$
will also be proportional to the string coupling:
\eqn\prop{\lambda_1\sim g^{\phantom{2}}_{\rm IB}\ell_s.}

Furthermore, the same operator will control the emission of closed
strings by an open string (see fig.~7(c)), because away from the open
string end--points, the operator is not sensitive to whether it is
splitting or joining a closed or an open string.

The operator ${\widetilde{\cal O}}$ will also control the open--open
string interaction which does not involve the end--points. This is the
case where two open string segments split and join somewhere along
their length to give two final open strings (see fig.~7(a)). This
interaction is again independent of whether the strings have
end--points.

\midinsert{
\vskip0.1cm
\centerline{\epsfxsize4.0in\epsfbox{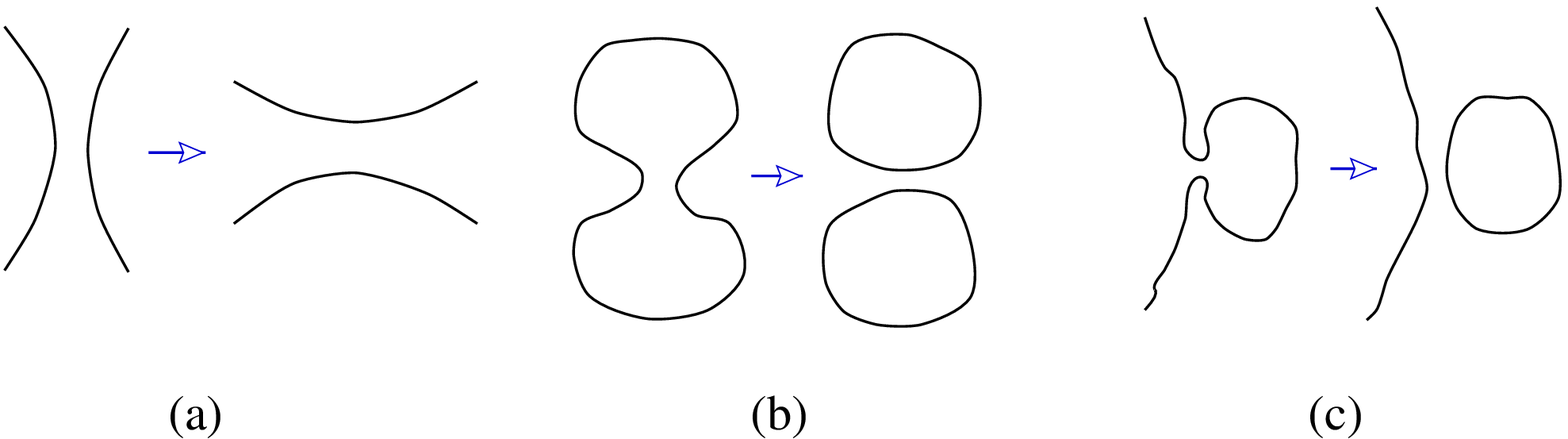}}
\vskip0.1cm
\centerline{\sl Figure 7.}}
\endinsert

Another type of interaction which the operator will control is the
orientation reversal of a string somewhere along its bulk (See
fig.~8). This is because of its symmetrical form.

\midinsert{
\vskip0.1cm
\centerline{\epsfxsize2.5in\epsfbox{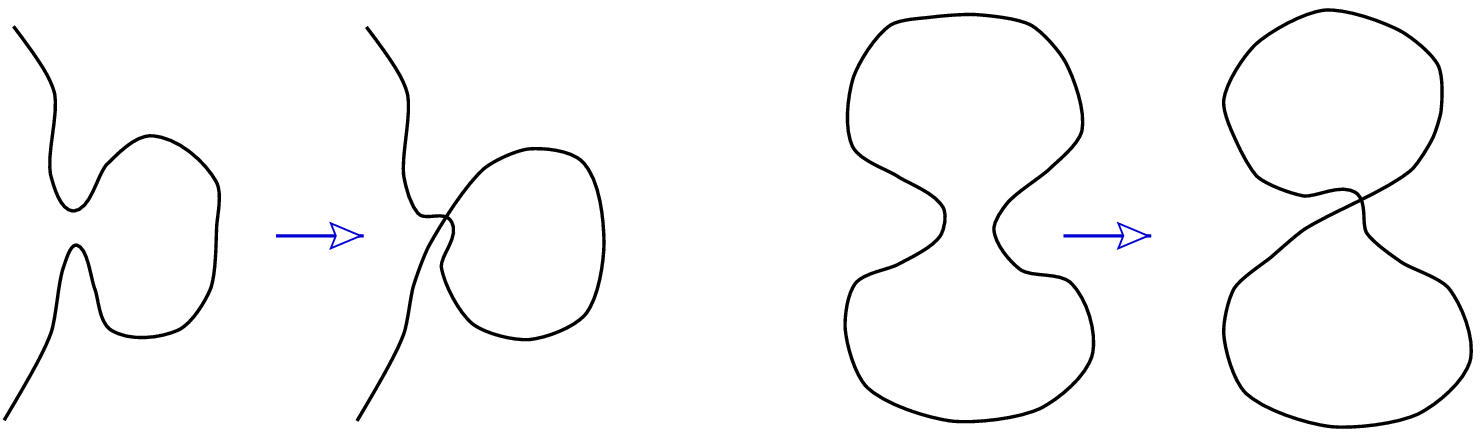}}
\vskip0.1cm
\centerline{\sl Figure 8.}}
\endinsert

\subsec{String End--Point Interactions}
The other string interaction we must consider is the case where
the end--points of open strings can join to form a longer open string, or a
closed string; or when a string (open or closed) can split somewhere
along its length to produce new free end--points (see fig.~9).

\midinsert{
\vskip0.1cm
\centerline{\epsfxsize2.5in\epsfbox{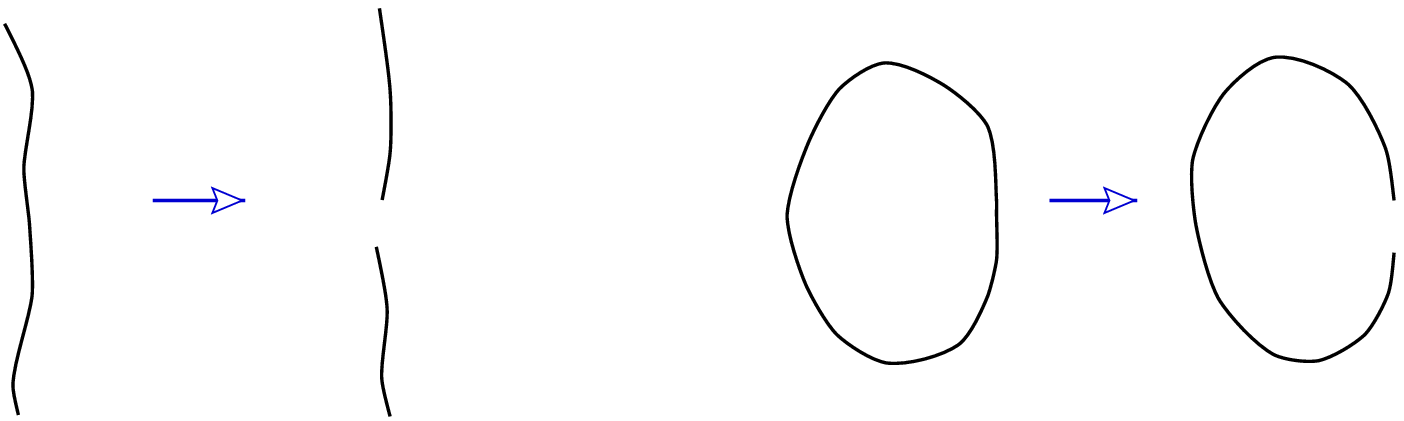}}
\vskip0.1cm
\centerline{\sl Figure 9.}}
\endinsert

As string end--points can only exist on the boundaries
$\sigma{=}0,\pi$, we can anticipate that the relevant conformal field
theory operator responsible for this interaction must only operate
there.

It is easy to see where this boundary operator comes from. We may
include an operator which obeys the reflection conditions in the
previous section by requiring that ${\bar{\cal O}}({\bar z})$ is the
{\it same} function of ${\bar z}$ as ${\cal O}(z)$ is of $z$. We have
effectively inserted a mirror on the boundary, selecting only a single
functional dependence for our operator ({\it i.e.} only the
holomorphic or anti--holomorphic part of ${\cal O}(z,{\bar z})$). So
this operator is naturally defined on the upper half plane (disc,
${\rm D}^2$), the full plane being a double cover.

We may infer the existence of boundary operators in the standard way,
examining the operator product expansion of ${\cal O}(z)$ with its
mirror image:
\eqn\mirrorope{{\cal O}(z){\cal O}({\bar z})\sim\sum_i 
{1\over(z-{\bar z})^{(3-h_i)}}{\cal B}^{(i)}(\tau)+\ldots} where $h_i$
are the wieghts of the operators $B^{(i)}$.  
A natural guess for the operator we need is the part
of ${\cal O}(z)$ living on the boundary:
\eqn\bound{{\cal B}_{ab}(\tau)=\lambda_{ab}
[\tau^i\Sigma^i](z)|_{\sigma=0,\pi}.}

(We have endowed the operator with a Chan--Paton matrix $\lambda_{ab}$
labeling which D8--branes lie on the boundary. This will in turn be
carried by the string end--points which are involved in the
interaction.)

This operator is added into the theory as follows (again summing over
all possible eigenvalues):
\eqn\endvertex{\lambda_2\int d\tau \sum_{I<J}{\cal B}^i_{IJ}(\tau).}

A quick way to visualize that this is the correct operator\foot{See
ref.\aki\ for another example of a twist operator on the boundary
introducing open string sectors.}\ is to think of this interaction as
a descendent of the oriented closed--closed interaction, where we have
subsequently identified the two halves of the closed string to make
open strings (see fig.~10). The splitting--joining interactions of the
parent type~II string also controls the splitting joining interaction
of end--points. The required operator $B(\tau)$ is simply the diagonal
action of $\widetilde{\cal O}$ at the fixed points of the projection
which descends open strings from closed ones. (In some sense, one can
think of the boundary operator as arising in the twisted sector of the
$\sigma{\to}-\sigma$ reflection operation, restricted to live at the
fixed points, $\sigma{=}0,\pi$. This is reminiscent of the
observations made in ref.\petrorientifold.)

\midinsert{
\vskip0.1cm
\centerline{\epsfxsize3.5in\epsfbox{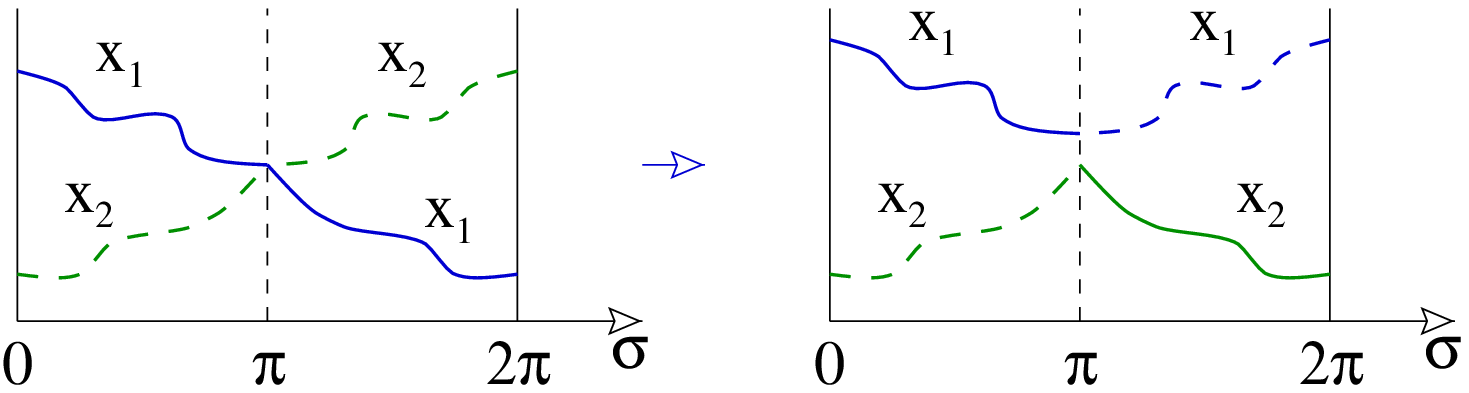}}
\vskip0.1cm
\centerline{\sl Figure 10}}
\endinsert

As ${\cal B}(\tau)$ has conformal dimension $3\over2$ and we are
integrating over the boundary, the coupling $\lambda_2$ must have
dimension $1\over2$. Evidently, it must be the square root of
$\lambda_1$, and so
\eqn\open{\lambda_2\sim g_{\rm IB}^{1\over2}\ell_s^{1\over2}.}

This is exactly what we want, for the splitting--joining interaction
of string end--points should have the Yang--Mills interaction
strength, which in turn should be the square root of the string bulk
interaction strengths, controlling gravity.

In other words, we have recovered the well--known relation
\eqn\wellknown{g^2_{\rm YM}\sim g^{\phantom{2}}_{\rm IB},}
familiar from D--brane physics\refs{\dbranes,\gojoe}. Here, of course,
the D--branes in question are type~IB's D9--branes, and the
Yang--Mills theory is $SO(32)$.

\newsec{Closing Remarks}
In summary, we have constructed the matrix theory (first arrived at in
ref.\super) of the weakly coupled second--quantized type~IB
string. The Fock space is built by ``winding'' type~IA strings.

The theory of the free second--quantized type~IB string theory is a
1+1 dimensional\foot{There is a different proposal in the
literature\morten.}\ orbifold conformal field theory which is
characterized by a matrix type~IB Green--Schwarz action, which has a
discrete Abelian gauge symmetry, the permutations of the
eigenvalues. 

The twisted sectors of the new theory contain long strings, which are
both open and closed. String end--points exist at the fixed points of
the type~IA orientifold group $\{1,\Omega{\cal R}_9\}$, which define
the type~IA interval. The D8--branes and orientifold O8--planes there
supply the Chan--Paton factors of the open strings, and $SO(32)$ gauge
symmetry arises by the correct assembly of the $SO(16){\times}SO(16)$
charged strings from the different twisted sectors.

There are two classes of string interactions in this open--closed
string theory, and we constructed them as operators in the orbifold
conformal field theory. 

The first class, which involves splitting and joining in the interior
of open and closed strings, has the strength of the closed string
coupling $g^{\phantom{2}}_{\rm IB}$, and is the non--orientable
descendant of the type~IIB interaction vertex, which is left--right
symmetric by construction.

The second class involves the creation or destruction of string
end--points, and is controlled by a boundary operator which is a
``descendant'' of the bulk interaction twist operator, in the sense
that it is the part of it restricted to the boundary. Appropriately,
it has the strength of the square root of the string coupling.

This completes the demonstration that all\super\ the five ten
dimensional superstring theories near weak coupling have a string
field theory description in terms of winding strings of the T--dual
species, where an economical description is given in terms of a matrix
Green--Schwarz Lagrangian.

Let us continue the discussion begun at the end of section~1.1.
It is not clear to the author whether the natural appearance of a
light--cone second--quantized field theory of strings from matrix
theory represents more than a delicious irony. 

Perhaps it is possible that such a second--quantized string
description will {\it always} appear naturally in the stringy limit of
{\it any} definition of M--theory.

This begs the question as to whether a possible route to a covariant
definition of M--theory may be sought by studying a (possibly
discretized) version of covariant string field theory. 

It may be that there are some important lessons to be learned from
covariant string field theory after all.

\bigskip
\bigskip

\noindent
{\bf Acknowledgments:}

\noindent
CVJ was supported in part by family, friends and music. This research
was supported financially by NSF grant
\#PHY97--22022.

\bigskip
\bigskip

\centerline{\epsfxsize1.0in\epsfbox{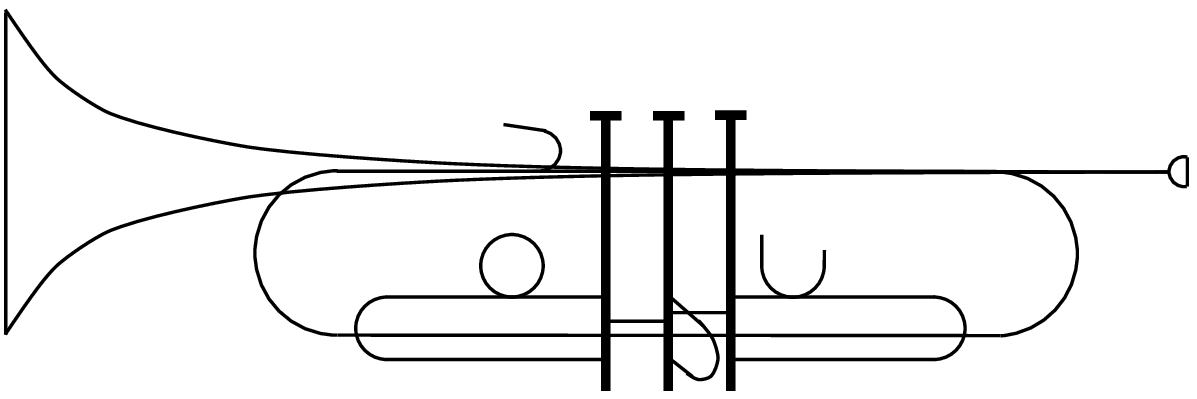}}

\listrefs

\bye